\documentclass[a4paper, 12pt]{article}
\usepackage[dvips]{graphicx}
\begin{document}

\normalsize
\begin{quote}
\raggedleft IFIC 99-28\\
\end{quote}
\vspace{0.5cm}
\begin{center}
\large {\bf $\sigma_{tot}^{pp}$ ESTIMATIONS AT VERY HIGH ENERGIES} \\
\bigskip
\vspace*{0.5cm}
J. Velasco$^{c}$,
J. P\'{e}rez-Peraza$^{a}$,
A. Gallegos-Cruz $^{b}$\\
M. Alvarez-Madrigal$^{a}$, 
A. Faus-Golfe$^{c}$, 
A. S\'{a}nchez-Hertz $^{a}$
\end{center}
\begin{quote}
{\small \it  
\hspace*{0.cm}$^{a}$Instituto de Geofisica, UNAM, 04510, C.U., Coyoacan, 
Mexico D.F.. MEXICO\\
\hspace*{0.cm}$^{b}$ Ciencias Basicas, UPIICSA, I.P.N., Te 950, Iztacalco,
08400, Mexico D.F, MEXICO \\
\hspace*{0.cm}$^{c}$IFIC, Centro Mixto CSIC-UV,
Dr. Moliner 50, 46100 Burjassot,
\hspace*{0.cm}Valencia, SPAIN}

{\small \it
\hspace*{0.cm}Presented at the 20th International  Cosmic Rays Conference,
Utah 1999}

{\small
\begin{center}
{\bf Abstract}
\end{center}
Proton-proton total cross sections ($\sigma_{tot}^{pp}$) are measured 
with present day
high energy colliders up to  2 TeV in the centre-of-mass of the
system ($ 10^{15}$ eV in the laboratory). Several
parameterizations, very succesful at low energies, can then be
used to extrapolate the measured values and get estimations of
cross sections to higher energies ($ 10^{17}$ eV).
On the other hand, from very high energetic cosmic rays
($\geq 10^{17}$ eV) and using some approximations, it is possible to
get a value for $\sigma_{tot}^{pp}$  from the knowledge 
of the $\sigma_{tot}^{p-air}$ at these energies.
Here we use a phenomenological model to estimate $\sigma_{tot}^{pp}$ 
at cosmic ray energies.
On the basis of regression analysis we show that the predictions 
are highly sensitive to the
employed data for extrapolation. Using data at 1.8 TeV 
our extrapolations for $\sigma_{tot}^{pp}$ 
are incompatible with most of cosmic ray results.}
\end{quote}
\vspace*{-0.5cm}
\subsection*{Hadronic $\sigma_{tot}^{pp}$ from accelerators and cosmic rays}

Since the first results of the Intersecting Storage Rings(ISR)
at CERN arrived in the 70s, it is a
well-established fact that $\sigma_{tot}^{pp}$  
rise with energy \cite{A1}, \cite{AME1}. The CERN $S\bar{p}pS$ Collider 
found this rising valid for $\sigma_{tot}^{\bar{p}p}$ as  well \cite{UA41}. 
Several parametrizations (purely theoretically, empirical
or semi-empirical based) fit pretty well the data. 
All of them agree that at the
energies (14 TeV in the centre-of-mass)
of the future CERN Large Hadron Collider (LHC) 
or higher the rise will continue. 
A thoroughful discussion on these problems may be found in \cite{Giorgio},
\cite{Blois97}.
\par
For our purposes  we have chosen a parametrization used 
by experimentalists to fit their
data \cite{UA42ext}. 
The most interesting piece
is the one controling the high-energy behaviour, 
given by a $ln^{2}(s)$ term, in
order to be compatible, asymptotically, with the Froissart-Martin bound
\cite{FM}.
The parametrization assumes 
$\sigma_{tot}^{pp}$ and $\sigma_{tot}^{\bar{p}p}$ to be the same
asymptotically. 
It has shown its validity   
predicting, from the ISR
data (23-63 GeV in the center of mass), 
the $\sigma_{tot}^{\bar{p}p}$ 
value found at the 
$S\bar{p}pS$ Collider (546 GeV), 
one order of magnitude higher in energy \cite{A2}, \cite{UA41}. 
\par
With the same well-known technique and using the most recent results
it is possible 
to get estimations for $\sigma_{tot}^{pp}$ at the energies
of the LHC and beyond \cite{UA42ext}. 
These estimations, together with our present experimental knowledge for both 
$\sigma_{tot}^{pp}$ 
and 
$\sigma_{tot}^{\bar{p}p}$ 
are summarized in Table 1 and plotted 
in fig. 1. We have also plotted the cosmic ray 
experimental data \cite{AKENO}, \cite{FLY1}.  The curve is the result 
of the fit describe in \cite{UA42ext}.
\par
The increase in $\sigma_{tot}$ as the energy increases 
is clearly seen.  
The main conclusion from this analysis based on accelerators
results are the predictions
$ \sigma_{tot}  = 109 \pm 8 $ mb at $\sqrt s = 14 $ TeV and   
$ \sigma_{tot}  = 130 \pm 13$ mb at $\sqrt s = 40 $ TeV.
\begin{table}[ht]
\centering
\begin{tabular}{|l||l||r|} \hline
$\sqrt s$ (TeV)& &$ \sigma_{tot}$ (mb) \\ \hline 
0.55 & Fit  & $61.8 \pm 0.7 $\\
     & UA4  & $62.2 \pm 1.5 $\\
     & CDF  & $61.5 \pm 1.0 $\\ \hline             
0.90 & Fit  & $67.5 \pm 1.3 $\\
     & UA5  & $65.3 \pm 1.7 $\\ \hline
1.8  & Fit  & $76.5 \pm 2.3 $\\
     & E710 & $72.8 \pm 3.1 $\\
     & CDF  & $80.6 \pm 2.3 $\\  \hline
14   & Fit  & $109 \pm 8 $   \\ \hline
40   & Fit  & $130 \pm 13 $  \\ \hline
\end{tabular} 
\caption{$\sigma_{tot}^{\bar{p}p}$ data from high-energy accelerators. Fits
values from  \cite{UA42ext}.}
\end{table}

\begin{figure}[ht]
\begin{center}
\mbox{\includegraphics[height=70mm,width=100mm]
{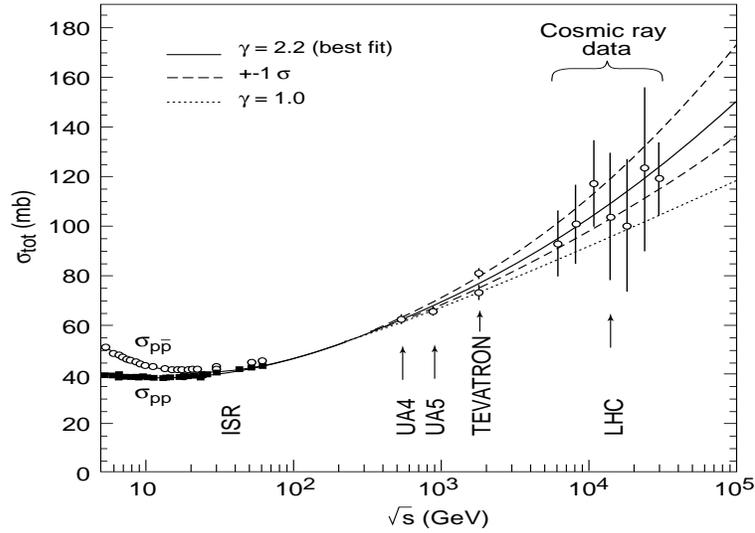}}
 \caption{ 
$\sigma_{tot}^{pp}$ and $\sigma_{tot}^{\bar{p}p}$ with the
prediction of \cite{UA42ext}.}
\end{center}
\end{figure}
Cosmic rays experiments give us  $\sigma_{tot}^{pp}$  
as derived from cosmic ray extensive air shower (EAS) data
 \cite{ECCR}. 
The primary interaction involved in EAS is  proton-air; 
what it is determined through EAS is 
the $p$-inelastic cross section, $\sigma_{inel}^{p-air}$. 
But the determination of $\sigma_{inel}^{p-air}$
(or its relation with $\sigma_{tot}^{pp}$) is model dependent.
A theory for nuclei interactions must be used.
Usually is Glauber's theory \cite{Glauber}.
The AKENO Collaboration has quoted, from  their results 
for a center-of-mass energy in the interval 6-25 TeV, 
$\sigma_{tot}^{pp}= 133 \pm 10$ mb at
$ \sqrt s = 40 $ TeV. On the other hand, an analysis  of 
the Fly's Eye experiment results
\cite{FLY2} claims $\sigma_{tot}^{pp}= 175_{-27}^{+40} $ mb 
at $\sqrt s = 40 $ TeV.
It has been argued by Nikolaev that this  contradiction in the values of 
both experiments disappears if, in the AKENO analysis,  
the $\sigma_{inel}^{p-air}$ is
identified with an absorption cross section \cite{Nik}. 
He obtains $\sigma_{tot}^{pp}= 160-170 $ mb 
at $\sqrt s = 40 $ TeV, which solves the discrepancy.
\vspace*{-0.5cm}
\subsection*{Are accelerators and cosmic ray $\sigma_{tot}^{pp}$ compatible?}
The results from cosmic ray experiments, from  
the previous analysis, have been
made compatible among themselves. 
But they have shifted away from
the estimations obtained with extrapolations using the data from 
accelerators. 
\par
The validity of these extrapolations, of course, may be discussed. 
But we would like to point
to the fact that  most extrapolations (as those using a 
$ln(s)$ term to control the
high-energy behaviour) predict even lower values 
for $\sigma_{tot}^{pp}$.  That makes the difference bigger.
\par
We have tackled the problem 
using the multiple-diffraction model 
\cite{Glauber}, \cite{GV1}. 
In a recent version of it \cite{MM1}  
the parameters of the model are determined fitting 
the $pp$ accelerator data 
in the interval $13.8 \leq \sqrt s
\leq 62.5 $ GeV.  The $\sigma_{tot}^{pp}$ values obtained when
extrapolated to higher energies seem to confirm the above quoted
compatible values of the cosmic ray experiments.  
That would imply the extrapolation cherished by experimentalists is wrong.
\par
But this approach predicts a value for $\sigma_{tot}^{pp}$ 
at the Fermilab
Collider (1.8 TeV) which seems to be very high: $91.6 $ mb 
(no error quoted). 
In table 1 we
see  that the measured $\sigma_{tot}^{\bar{p}p}$ at that energy 
is much smaller. It may be argued that $\sigma_{tot}^{pp}$  and
$\sigma_{tot}^{\bar{p}p}$ are different at high energies.
This is the ``Odderon hypothesis'', which
has been very much weakened recently  \cite{UA42}.
\par
Taking this into account, in our 
multiple-diffraction analysis 
it is assumed  the same behaviour
for $\sigma_{tot}^{pp}$ and $\sigma_{tot}^{\bar{p}p}$ at high energy.  
Results are summarized in table 2 and  plotted in fig. 2.  
\vspace*{0.3cm}
\begin{center}
\begin{tabular}{|l||l||l||l||l||l||r|} \hline
$\sqrt s$ (TeV)& $ \sigma_{tot}$ (mb)& $ \sigma_{upp}$ (mb)& $ \sigma_{low}$ (mb)&  
$ \sigma_{tot}$ (mb)& $ \sigma_{upp}$ (mb)& $ \sigma_{low}$ (mb)\\ \hline
0.55  & 69.39  & 77.77  &  62.0   & 62.24  & 63.56  & 60.98   \\ \hline 
0.9   & 78.04  & 89.62  &  67.87  & 67.94  & 69.35  & 66.59   \\ \hline      
1.8   & 91.74  & 108.64 &  76.99  & 76.44  & 78.14  & 74.84   \\ \hline 
14    & 143.86 & 182.45 &  110.32 & 104.17 & 108.57 & 99.85   \\ \hline 
40    & 177.23 & 230.32 &  130.95 & 118.99 & 125.98 & 111.75  \\ \hline
\multicolumn{4}{c}{\hspace{1.8cm}(a)}               & \multicolumn{3}{c}{(b)}      \\ 
\end{tabular}
\vspace*{0.3cm}
\\Table 2:  Predicted $\sigma_{tot}^{pp}$ from fitting 
accelerator data at \\ (a) $\sqrt s \leq $ 62.5 GeV; 
(b) including 
data at 546 GeV and 1.8 TeV.
\end{center}

\begin{figure}[ht]
\begin{center}
\mbox{\includegraphics[height=95mm,width=160mm]
{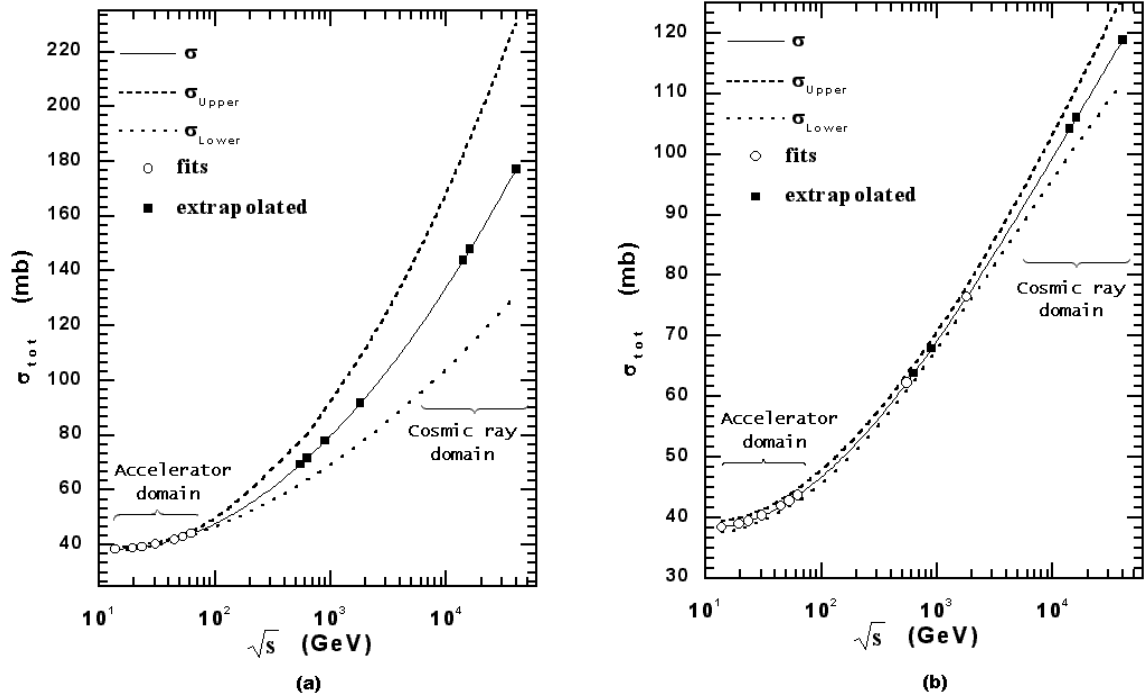}}
\caption{Predictions (black squares) of $\sigma_{tot}^{pp}$:  
(a) data at $\sqrt s \leq $ 62.5 GeV;
(b) including data at 546 GeV and 1.8 TeV (open circles).}
\end{center}
\end{figure}
\par
Our results indicate that, if in the phenomenological 
multiple-diffraction approach we limit
our fitting calculations to the accelerator domain
$\sqrt s \leq 62.5 $ GeV, the extrapolation to high 
energies is in complete agreement with the analysis carried out by Nikolaev
\cite{Nik}, and with the experimental data of the Fly's Eye  
\cite{FLY2} and the Akeno \cite{AKENO}
collaborations, because their quoted 
errors fall within the error band of our extrapolations. That is, such an 
extrapolation produces an error band so large at cosmic ray energies that any
cosmic ray results become compatible with results at accelerator energies.
However, if additional data at higher accelerator energies 
are included and then the error band obviously narrows, things change.
This can be 
seen in fig.2b, where we have considered data at 0.546 TeV and 1.8 TeV 
(according to Table 1),
in which case the predicted values of $\sigma_{tot}^{pp}$ 
from our extrapolation at $\sqrt s = 40 $ TeV, 
$\sigma_{tot}^{pp} = 119 \pm 7 $ mb
are much lower than those illustrated in fig. 2a, 
and clearly incompatible 
with the reinterpreted Fly's Eyes and Akeno results by several standard
deviations.
Concerning the quoted error bands 
we employed the so called 
forecasting technique of regression analysis \cite{regres}.
\par
We conclude that, when all experimental  available data is taking into
account, the estimated values for $\sigma_{tot}^{pp}$ obtained from 
extrapolation from present high-energy accelerators and those obtained from
cosmic ray experiments are incompatible in the region around 
$\sqrt s = 40 $ TeV mb. 

\end{document}